\documentclass{new_tlp}
\pdfoutput=1
\usepackage{epsfig}
\usepackage{makeidx}  
\usepackage{amsmath}
\usepackage{draftwatermark}
\SetWatermarkText{DRAFT}
\SetWatermarkScale{1}

\newcounter{example}
\newenvironment{example}[1][]{\refstepcounter{example}\par\medskip
   \noindent \textbf{Example~\theexample. #1} \rmfamily}{\medskip}
\newtheorem{dummylemma}{Dummy}[section]
\newtheorem{DefinitioN}[dummylemma]{\bf Definition}
\newenvironment{definition}[1]
               {\begin{DefinitioN}\rm {\bf (#1)}}{\hspace{1cm} \hfill \end{DefinitioN}}

\newtheorem{theorem}{Theorem}
\newtheorem{lemma}{Lemma}[section]

\title[Theory and Practice of Logic Programming]
      {The Semantics of Metaprogramming in Prolog}
\author[D. S. Warren]
       {DAVID S. WARREN\\
         Stony Brook University, Stony Brook, NY, 11794-4400 USA\\
         \email{warren@cs.stonybrook.edu}\\
         XSB Inc., Setauket, NY, 11733 USA\\
         \email{warren@xsb.com} }
\begin{document}
\maketitle              
\begin{abstract}
  This paper describes a semantics for pure Prolog programs with negation that provides meaning to metaprograms.  Metaprograms are programs that construct and use data structures as programs.  In Prolog a primary mataprogramming construct is the use of a variable as a literal in the body of a clause.  The traditional Prolog 3-line metainterpreter is another example of a metaprogram.  The account given here also supplies a meaning for clauses that have a variable as head, even though most Prolog systems do not support such clauses.  This semantics naturally includes such programs, giving them their intuitive meaning.

  Ideas from M. Denecker and his colleagues \cite{DBLP:journals/tocl/DeneckerBM01,DBLP:journals/corr/abs-2304-13430} form the basis of this approach.  The key idea is to notice that if we give meanings to all propositional programs and treat Prolog rules with variables as the set of their ground instances, then we can give meanings to all programs.  We must treat Prolog rules (which may be metarules) as templates for generating ground propositional rules, and not as first-order formulas, which they may not be.  We use parameterized inductive definitions to give propositional models to Prolog programs, in which the propositions are expressions.  Then the set of expressions of a propositional model determine a first-order Herbrand Model, providing a first-order logical semantics for all (pure) Prolog programs, including metaprograms.

We give examples to show the applicability of this theory.  We also demonstrate how this theory makes proofs of some important properties of metaprograms very straightforward.
\end{abstract}

\section{Introduction}

The simplest use of metaprogramming in Prolog is in the definition of the standard predicate {\tt call/1}, which is defined with the single rule:
\begin{verbatim}
call(X) :- X.
\end{verbatim}
This allows a Prolog programmer to construct a term at runtime into a variable, say {\tt Goal}, then to use that data structure as a query to the Prolog engine by invoking {\tt call(Goal)}.  In most Prolog implementations the programmer could simply use {\tt Goal} itself instead of {\tt call(Goal)}. 

Another well-known example of a Prolog metaprogram is the 3-line metainterpreter:
\begin{verbatim}
  interp(true).
  interp((A,B)) :- interp(A), interp(B).
  interp(H) :- rule(H,Body), interp(Body).
\end{verbatim}
Here {\tt rule/2} returns a ``program component,'' i.e., a rule.  This is another example of a metaprogram, in which data, the facts of the predicate {\tt rule/2}, are treated as a program.  Prolog systems normally allow programmers to retrieve the rules of the executing program by using a built-in predicate {\tt clause/2}. By using {\tt clause/2} the Prolog programmer has direct access to the code (rules) of the executing program.

Now consider the following ``Prolog'' metaprogram:
\begin{verbatim}
  true.
  (A,B) :- A, B.
  H :- clause(H,Body), Body.
\end{verbatim}
The program is not an ISO standard Prolog program.  The first two clauses are (potentially) fine; it's the third, with a variable as the rule head, that is problematical.  But if we use our Prolog understanding, it is easy to see how this program ought to work.  We can simply wrap every literal in this program with unary predicate symbol {\tt interp/1} and we get (almost) the vanilla metainterpreter above, in which there is no restriction on the first argument of {\tt rule/2} facts.  (Note that the call to {\tt clause/2} remains unwrapped.)

We will assume in this paper that our Prolog programs do allow variables in heads of rules.  And so this is a valid Prolog program for us.

The usual semantics for Prolog is based on a logical theory: the least Herbrand Model satisfying all its Horn clauses \cite{DBLP:journals/jacm/EmdenK76,DBLP:journals/jlp/Fitting85}, or a model satisfying the rules as inductive definitions of relations \cite{DBLP:journals/tocl/DeneckerBM01,DBLP:journals/corr/abs-2304-13430}.  These semantics do not account for metaprograms with variables in bodies or as heads of rules.  This is because heads and body atoms have been treated as atomic formulas of a first-order logic, and a variable is {\em not} an atomic formula.  Thus, these formulations fail for such metaprograms.  And even though they can provide a meaning for our {\tt interp/1} program above, there is no clear semantic connection between the data structures returned by {\tt rule/2} and the program that is metainterpreted.  Our semantics makes these connections and, to do so, must give more refined meanings to Prolog programs.

Rather than mapping the head and body literals to atomic formulas in first-order logic, or assertions of membership in relations, we consider them as propositional constants, requiring they be ground.  If a rule has variables, it is treated as a template for all its ground instantiations.  And this is where metarules, i.e., those not built from first-order atomic formulas, become sets of valid propositional formulas, i.e., ground rules.  Note that these propositions are structured, i.e., they are expressions, a.k.a., ground atomic formulas or ground terms.  A propositional model of a set of propositional formulas is a set of propositions, in our case, a set of ground atomic formulas.  And a set of ground atomic formulas uniquely determines a first-order Herbrand structure.  In this way, we give a first-order model theoretic meaning to a Prolog (meta-)program.

To give meaning to ground programs, we use parameterized inductive definitions as proposed and developed by Denecker et al. \cite{DBLP:journals/tocl/DeneckerBM01}.  This provides logical meanings for rules as inductive definitions and allows programs to be made up as the combination of separate components, each with an independent meaning.  Thus, it provides a compositional semantics for Prolog programs.   A component traditionally must contain all rules with the same head predicate symbol.  In our propositional case, we have no obvious relations.  We use instead any set of rules in the program and provide conditions on the set of their heads under which such rule sets can be a component.

The body of this paper first develops the semantics for positive programs, providing several equivalent characterizations of the set of propositions determined by interpreting a set of propositional rules as a parameterized inductive interpretation.  It then discusses how parameterized components can be combined to construct larger components and ultimately full programs.  Then the theory of parameterized components is extended to rules with negation, thereby providing a semantics for stratified programs.  We omit proofs from this paper both due to space limitations and also because many of them are straightforward.  The paper concludes with a discussion of the implications of this theory for metaprogramming, and for the interpretation of metaprograms in first-order logic.

\section{Inductive Definitions of Sets of Propositions}

We all have encountered many inductively defined sets in our journeys through mathematics.  The first example might be the set, $Nat$, of natural numbers, i.e., 0, 1, 2, ..., whose definition can be: \[
x \in Nat = \left\{ \begin{array}{cl}
  \mbox{if} & x = 0 \\
  \mbox{if} & x' \in Nat \mbox{ and } x = succ(x') 
\end{array}\right.
\]
meaning that $x$ is a natural number if it is $0$, or if it is the successor of a natural number. 

Another example is the set $R_a^G$ of nodes reachable in a graph $G$ from node $a$:
\[
x \in R_a^G = \left\{ \begin{array}{cl}
  \mbox{if} & x = a \\
  \mbox{if} & y \in R_a^G \mbox{ and there is an edge in $G$ from }y \mbox{ to } x
\end{array}\right.
\]

As a simple example of a definition (perhaps not inductive but still rule-based) is the set of university requirements that a student has satisfied on their journey toward a degree.  We might define (schematically for a hypothetical university):
\begin{verbatim}
met_cs_calc_reqs if took_calc_I and took_calc_II 
met_cs_calc_reqs if took_calc_A and took_calc_B and took_calc_C
met_cs_math_reqs if met_cs_cals_reqs and took_discrete_math
met_cs_intro_pgming_reqs if took_pgming_I and took_pgming_II
met_cs_adv_pgming_reqs if met_cs_intro_pgming_reqs and took_algorithms
...
met_graduation_reqs if met_cs_reqs and met_distribution_reqs
\end{verbatim}
We have left out many rules but expect the reader can envision how they could be completed.  The idea is that if we plug in a student's transcript to define what {\tt took\_*} propositions hold, then these rules define the set of requirements that that student has met, the goal of most students being to generate a transcript that causes {\tt met\_graduation\_reqs} to show up in the set defined by these rules.

These rules can be understood as rules in a propositional logic: all the symbols in these rules (except the {\tt if} and {\tt and}) are propositional constants, or propositions.  The set of requirements met by a particular student is an inductively defined set of propositions generated from those rules and facts.  So, our goal is to precisely define propositional models for sets of rules and facts.

Thus, we turn to a general theory behind inductive definitions of sets of propositions.  An inductively defined set is determined by a collection of propositional rules on a set of propositions.  We let $U$ denote the set of all propositions.

We represent a propositional rule in positive Prolog:
\[ p_0 \mbox{ :- } p_1, p_2, ..., p_n \]
as the slightly more abstract pair $(p_0, \{p_1, p_2, ..., p_n\})$, turning the rule body sequence into a set.  The general form of a propositional Prolog rule will be $(h,B)$ where $h \in U$ and $B \subseteq U$.  We call such a pair simply a rule.  For rule $(h,B)$, $B$ is the {\em body} of the rule and $h$ is the {\em head}.


We note that the propositions in $U$ may be structured.  To handle grounded Prolog rules, we take $U$ to be the set of all expressions (or trees) over finite string symbols.  This set is the set of all legal Prolog ground terms.  In logic, this set of trees is both the set of ground atomic formulas, and the set of ground terms.  We will generally use the neutral word ``expression'' for a member of this set when we want to stress its structure; otherwise, we may call it a proposition.  $U$ will be the set of expressions throughout our development.

\begin{definition}{Inductive Definition}
  An {\em inductive definition} ${\cal P}$ on set $U$ is a set of rules on $U$.
\end{definition}
We use inductive definitions on the set of propositions $U$ to define sets of propositions.  Note that since for us $U$ is the set of expressions, an inductive definition will define a set of expressions, i.e., a set of atomic formulas.  And a set of atomic formulas uniquely determines a first-order Herbrand structure, that structure that is true of all and only atomic formulas in the set.  In this way we can think of our inductive definitions as determining first-order (Herbrand) models.  We will return to discuss this idea later.

\begin{definition}{A Set Closed under a Rule}
For rule $(h,B)$, a set $S$ is {\em closed under $(h,B)$} if  $h \in S$ when $B \subseteq S$.
\end{definition}

\begin{definition}{Set Containing $A$ and Closed under ${\cal P}$}
For inductive definition ${\cal P}$ on $U$, and set $A \subseteq U$, a subset $S$ of $U$ {\em contains $A$ and is closed under ${\cal P}$} if $A \subseteq S$ and $S$ is closed under every rule $R \in {\cal P}$.
\end{definition}
The same set of rules ${\cal P}$ can determine different closed sets depending on the set $A$, hence we call $A$ a {\em parameter set} for the rules ${\cal P}$.


\begin{lemma}
The intersection of two sets containing $A$ and closed under ${\cal P}$ also contains $A$ and is closed under ${\cal P}$.
\end{lemma}

\begin{definition}{Inductively Defined Set $F_{\cal P}(A)$}
The set determined by an inductive definition ${\cal P}$ on $U$ with parameter set $A$, $F_{\cal P}(A)$, is the least subset of $U$ containing $A$ and closed under ${\cal P}$.
\end{definition}

The previous lemma ensures that there is a unique minimal such set.  Thus, $F_{\cal P}(A)$ is well-defined.

\begin{definition}{Head and Body Sets of ${\cal P}$}
  The {\em Head Set, $H_{\cal P}$, and Body Set, $B_{\cal P}$, of rule set ${\cal P}$} are:
  \[H_{\cal P} = \{h: (h,B) \in {\cal P}\}\]
  \[B_{\cal P} = \bigcup_{(h,B)\in {\cal P}} B \]
\end{definition}
$H_{\cal P}$ is the set of propositions that appear in the head of any rule in ${\cal P}$.  Note that the least set $F_{\cal P}(A)$ will contain {\em only} elements from $H_{\cal P}$ or $A$.  I.e., $F_{\cal P}(A) \subseteq H_{\cal P} \cup A$ for any $A \subseteq U$.  In Prolog we usually think of a set of clauses as defining a predicate, and all rules with that predicate in the head as determining a specific subset of that predicate.  In our propositional framework, we use $H_{\cal P}$ to define the set from which the rules define a subset.  And since parameter sets for rules ${\cal P}$ should influence how ${\cal P}$ determines a set and not add elements directly, we require throughout that parameter sets to be {\em allowable}:

\begin{definition}{Allowable Parameters}
  Given an inductive definition ${\cal P}$ over $U$ and a set $A \subseteq U$, $A$ is an {\em allowable} parameter set for ${\cal P}$ if $A \cap H_{\cal P} = \emptyset$.  We require rule sets to have only allowable parameter sets.
\end{definition}

\begin{definition}{Propositional Model Satisfying ${\cal P}$}
A propositional model $M$ is a set of propositions, $M \subseteq U$.
\[ M \models {\cal P} \mbox{ if } F_{\cal P}((B_{\cal P} - H_{\cal P}) \cap M) = (H_{\cal P} \cap M) \] \end{definition}
The intuition behind this definition is the following.  The model $M$ itself provides the definition of the set $A$, a parameter set for ${\cal P}$.  That is, $A = (B_{\cal P} - H_{\cal P}) \cap M$.  Then given this parameter set, $M$ agrees precisely with the least set determined by it and ${\cal P}$ for all the head propositions of ${\cal P}$.  All other propositions of $U$ may or may not be in $M$.

\begin{figure}
  \includegraphics[width=20cm,height=8cm, trim=2.0cm 4.0cm 0.0cm 2.0cm,
    clip=true]{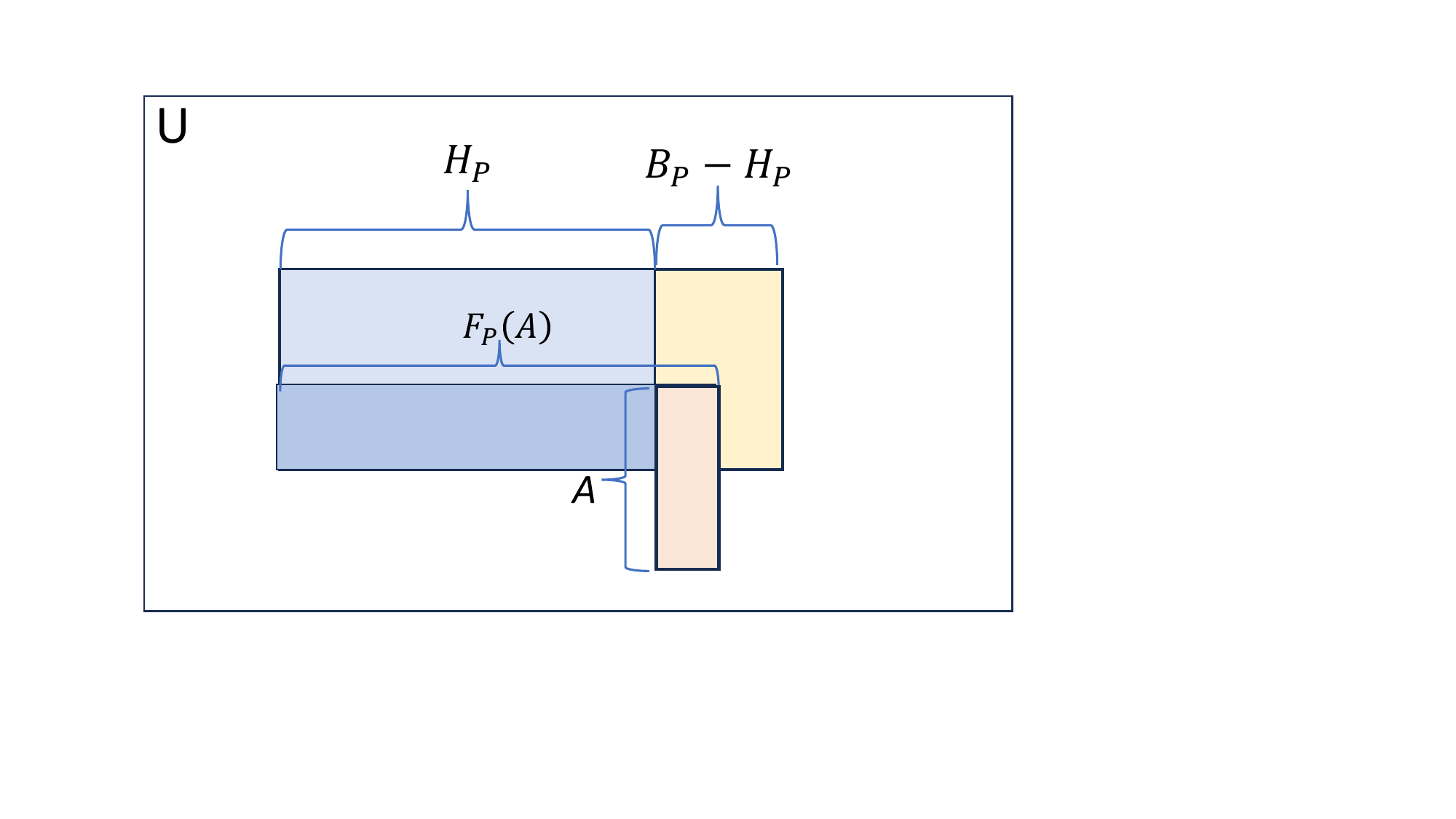}
  \caption{Venn Diagram for sets determined by rules ${\cal P}$}
  \label{simple-venn}
\end{figure}

We visualize the set of propositions of $U$ that appear in a set of rules ${\cal P}$ as shown in Figure \ref{simple-venn}. The outer box encloses all the propositions of $U$. The horizontal blue box contains the propositions in heads of rules, $H_{\cal P}$, and the yellow box represents the set of body propositions that are not also head propositions, $B_{\cal P} - H_{\cal P}$.  The vertical salmon-colored box represents an allowed parameter set $A$.  Note it is disjoint from $H_{\cal P}$.  The darker blue sub-box of $H_{\cal P}$ and the salmon box represent $F_{\cal P}(A)$.

We have characterized the set $F_{\cal P}(A)$ as the least set containing $A$ and closed under ${\cal P}$.  There are other ways to define this same set, and we turn to those other possible definitions.

\begin{definition}{The $T_{{\cal P},A}$ Set Function}
Given an inductive definition ${\cal P}$ on $U$ and parameter set $A \subseteq U$ and set $S \subseteq U$, let $T_{{\cal P},A}(S) = A \cup \{h : \exists (h,B) \in {\cal P} \mbox{ such that } B \subseteq S\}$.
\end{definition}
$T_{{\cal P},A}$ is a function on subsets of $U$ that simply accumulates all the heads of rules whose body elements are all in the input set and adds all elements of $A$.  $F_{\cal P}(A)$, the least set containing $A$ and closed under ${\cal P}$, can be defined in terms of $T_{{\cal P},A}$.

\begin{theorem} {\normalfont
  $F_{\cal P}(A)$ is the least fixpoint of the $T_{{\cal P},A}$ operator.   I.e., $T_{{\cal P},A}(F_{\cal P}(A)) = F_{\cal P}(A)$, and no proper subset of  $F_{\cal P}(A)$ has this property.
} \end{theorem}
We note that since $T_{{\cal P},A}$ is monotone operator on sets, it is known to have a least fixpoint.  This theorem claims that least fixpoint is $F_{\cal P}(A)$.

We now have two definitions of $F_{\cal P}(A)$, but neither presents any obvious way to compute membership in this set.  We next present two more definitions of $F_{\cal P}(A)$, which do lead to such algorithms.

\begin{definition}{The ${\bf T_{{\cal P},A}^{(i)}}$ Set Operator}
Let $T_{{\cal P},A}^{(0)} = \emptyset$; and $T_{{\cal P},A}^{(i+1)}$ = $T_{{\cal P},A}(T_{{\cal P},A}^{(i)})$.  Thus, $T_{{\cal P},A}^{(i)}$ starts with the empty set and iteratively applies $T_{{\cal P},A}$ to it $i$ times.
\end{definition}

\begin{theorem} {\normalfont
    $F_{\cal P}(A) = \bigcup_{i=0}^\infty T_{{\cal P},A}^{(i)}$.
} \end{theorem}

This countable union indeed defines $F_{\cal P}(A)$, the least set containing $A$ and closed under ${\cal P}$.  I.e., if we start with the empty set and iteratively apply the $T_{{\cal P},A}$ operator countably many times (or until nothing new is obtained), we obtain the least set of propositions containing $A$ and closed under the rules ${\cal P}$.

Since $T_{{\cal P},A}$ is monotonic, this is a non-decreasing sequence of sets.  Our universal set $U$ is infinite, so this might be an infinite sequence of ever-increasing sets and the defined set would be infinite.  If $H_{\cal P}$ is finite, then $F_{\cal P}(A)$ will be finite and for some natural number $n$, $F_{\cal P}(A) = \bigcup_{i=0}^n T_{\cal P}^{(i)}$.  A direct implementation of this construction provides a bottom-up algorithm for producing finite $F_{\cal P}(A)$.

Indeed there is yet another definition of $F_{\cal P}(A)$ that suggests another way to compute its members.  The iterated $T_{{\cal P},A}$ algorithm computes the entire proposition set $F_{\cal P}(A)$.  This next definition allows us, in some cases, to compute membership (and non-membership) in $F_{\cal P}(A)$ even when it is infinite.

\begin{definition}{Justification Sequence}
  A justification sequence for rules ${\cal P}$ on $U$ and parameter set $A$ is a finite sequence of propositions of $U$, $[p_0, p_1, p_2, p_3, ..., p_n]$ such that for every $p_i$ in the sequence, either $p_i \in A$ or there is a rule $(p_i,B) \in {\cal P}$ such that $B \subseteq \{p_0, p_1, ... p_{i-1}\}$.  I.e., every proposition of the sequence is either in $A$ or is the head proposition of a rule all of whose body propositions appear earlier in the sequence.
\end{definition}

\begin{theorem} {\normalfont
    $F_{\cal P}(A) = \{p_n : p_n$ is the final proposition of a justification sequence for ${\cal P}$ and $A\}$.
} \end{theorem}

With this characterization of $F_{\cal P}(A)$, we can show membership of a proposition of $F_{\cal P}(A)$ by producing a justification sequence for it.  A justification sequence for a given $p \in U$ can be constructed by first testing whether $p \in A$ and if not, finding a rule $(p,B)$ and then recursively constructing a justification sequence for each $b \in B$.  This produces a top-down (partial) algorithm for determining membership in $F_{\cal P}(A)$.  A top-down tabling algorithm also constructs a justification sequence but guarantees in addition that no proposition appears in it more than once.

\section{Parameters of Inductive Definitions and Program Components}

Thus far, we have thought of parameter sets as being directly provided.  But we can use a set of rules to define the parameter set for another set of rules.

\begin{example} {\normalfont
Consider again our example of university requirements.  Where requirement symbols of the form {\tt met\_*\_reqs} are defined in terms of other such symbols and transcript symbols of the form {\tt took\_*}.  The transcript symbols are all defined with facts, i.e., rules in which the body is the empty set.  For every student the requirement rules are the same, only the transcript rules differ.  Thus, we can naturally keep only the transcript rules in our set ${\cal P}$ rules, and let $A_{Std\_ID}$ be the set of transcript rules for the student with ID {\em Std\_ID}.  Then for any student with ID {\em X}, we can check whether $\mbox{met\_graduation\_reqs} \in F_{\cal P}(A_X)$ in order to know whether that student can graduate.
} \end{example}

\begin{example} {\normalfont
As another example of a parameterized inductive definition, consider transitive closure of a graph.  We will represent transitive closure facts with expressions of the form {\tt tc(X,Y)} where {\tt X} and {\tt Y} are integers representing nodes in a graph.  And we will represent edge facts with expressions of the form {\tt edge(X,Y)} where {\tt X} and {\tt Y} are integers representing nodes.  So, we can represent the transitive closure of a graph with nodes with the following rules ${\cal P}$:
\begin{verbatim}
tc(X,Y) :- edge(X,Y).
tc(X,Z) :- edge(X,Z), tc(Z,Y).
\end{verbatim}
where propositional rules are obtained by replacing {\tt X}, {\tt Y}, and {\tt Z} in all ways consistently by integers.  Thus, these rule templates generate propositional rules with propositions being expressions.  We can take $A$ to be some subset of propositions {\tt edge(X,Y)} (for {\tt X} and {\tt Y} integers).  With this setup, we can determine whether {\tt (I,J)} is in the transitive closure of the graph represented by the edges in $A$ by testing whether proposition {\tt tc(I,J)} is in $F_{\cal P}(A)$.  So, these two rule templates generate propositional rules that define the meaning of transitive closure.  Then for any graph of interest represented by $A$, these rules will give us the correct pairs in the transitive closure of the graph.
} \end{example}

\subsection{Nested Parameterized Inductive Definitions}

Consider how we might use one parameterized inductive definition to determine a parameter set for another parameterized inductive definition.  We can think of examples where this might be useful.

\begin{example} {\normalfont
    Consider again our transitive closure example:
\begin{verbatim}
tc(X,Y) :- edge(X,Y).
tc(X,Y) :- edge(X,Z), tc(Z,Y).
\end{verbatim}
Let the corresponding set of propositional rules be ${\cal TC}$.  Given a set $E$ of expressions of the form {\tt edge(A,B)} for some constants {\tt A} and {\tt B}, we have that $F_{\cal TC}(E)$ is the transitive closure of the graph $E$.  But perhaps our graph is over states, and the edges represent pairs of states in which the second is directly accessible from the first through some action.  We want to write a definition of this set of {\tt edge} pairs.  So, we write a definition, that might look something like:
\begin{verbatim}
edge(X,Y) :- applicable_action(X,Op), apply_action(Op,X,Y).
\end{verbatim}
where our edge definition checks and applies applicable actions that can move from one state to another.  Let the ground rules of this program be ${\cal MV}$.

Now we want the transitive closure of this particular {\tt edge} relation, the one defined by ${\cal MV}$.  We use ${\cal MV}$ to determine a set of edge expressions, and then use that set as the parameter set to the ${\cal TC}$ program.  And, assuming ${\cal MV}$ doesn't need a parameter set, this is exactly the set $F_{\cal TC}(F_{\cal MV}(\emptyset))$.
} \end{example}

This idea of using one component to generate parameter sets for another allows us to collect various program components and put them together.  There are clearly some constraints on the forms of rules sets that can be combined in this way.  The following theorem clarifies these constraints.

\begin{theorem} {\normalfont
Let ${\cal P}$ and ${\cal Q}$ be parameterized inductive definitions over $U$.  (We want ${\cal Q}$ to generate the parameter set for ${\cal P}$.)  Assume $H_{\cal P} \cap (H_{\cal Q} \cup B_{\cal Q}) = \emptyset$, i.e., the head set of ${\cal P}$ does not intersect with either the head set or the body set of ${\cal Q}$.  Then $F_{{\cal P} \cup {\cal Q}}(A) = F_{\cal P}(F_{\cal Q}(A))$.  I.e., the meaning of the union of the clause sets is the indicated composition, with ${\cal Q}$ providing the parameter set for use by ${\cal P}$.
} \end{theorem}

This constraint on the use of propositions of $U$ is clearly important.  Here ${\cal P}$  uses ${\cal Q}$ to generate its parameter set.  ${\cal Q}$ cannot define or use any proposition that is potentially defined by ${\cal P}$; it can define only propositions used by ${\cal P}$, i.e.,  propositions in $B_{\cal P} - H_{\cal P}$.  The set ($F_{\cal Q}(A))$ (of the clause set of the component below) may, and to be helpful does, intersect with the body set of the clause set of the component above.  (The body sets of the two components intersecting just suggests that they both might use yet another clause set as a parameter.)

\begin{figure}
  \includegraphics[width=20cm,height=8cm, trim=2.0cm 3.0cm 0.0cm 2.5cm,
    clip=true]{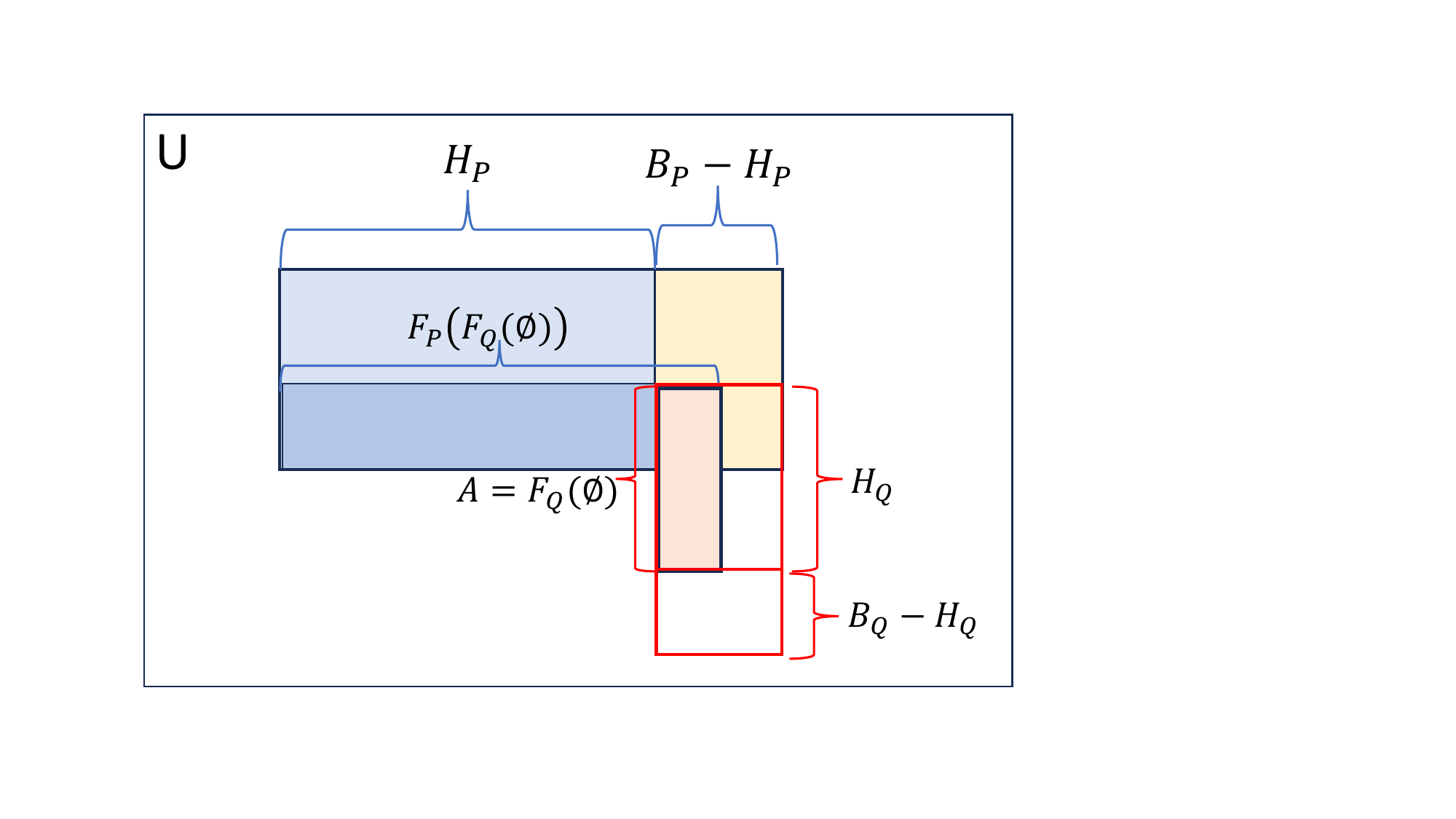}
  \caption{Venn Diagram for sets determined by parameterized ${\cal P}$}
  \label{theory-venn-par}
\end{figure}

Figure \ref{theory-venn-par} shows a Venn diagram for the propositions of a parameterized inductive definition ${\cal P}$ with parameter set $A$, which is generated as the least closed set of definition ${\cal Q}$.  I.e., component ${\cal Q}$ generates the parameter set for component ${\cal P}$.  The horizontal blue and yellow rectangle represents the propositions of ${\cal P}$, as indicated.  The vertical red-outlined rectangle represents the propositions of ${\cal Q}$.  Its head propositions overlap with ${\cal P}$'s body (but not head) propositions.  The smaller salmon-colored rectangle represents $F_{\cal Q}(\emptyset)$, as would be computed from ${\cal Q}$.  This set is a subset of $H_{\cal Q}$, disjoint from $H_{\cal P}$, intersects with $B_{\cal P} - H_{\cal P}$, and constitutes the parameter set $A$ for ${\cal P}$.

In this way we can build inductive definitions by using definitions to determine parameters to other parameterized definitions, as long as their head and body sets obey the necessary constraints.  We can build large definitions by using smaller components by combining them in this way.  Similarly, we may be able to decompose a large definition into subcomponents, again subject to the head and body constraints.  Note that these constraints will require that a set of mutually recursive definitions must all reside in the same component.  Thus, we have a compositional framework for parameterized inductive definitions.

\section{Inductive Definitions with Negation}

Prolog programs may have bodies with negative conditions, and we want to provide a semantics for such programs.  However, unconstrained use of negative conditions in cycles is known to lead to complications; traditional Prolog systems go into infinite loops on programs with those cyclic uses of negation.  It is known that Prolog meaningfully evaluates the so-called {\em stratified} programs, so this is a class of programs for which we want to provide a meaning here.

We note that the (allowable) parameter set for a parameterized inductive definition does not change in the bottom-up construction of the least model of the definition.  I.e., it can be constructed for a subcomponent independently, and then just used in the construction of the higher component.  So, we can allow a test for non-membership in a parameter set assuming we can ensure that this non-membership test will not change in the least model construction.  And we can ensure this by requiring that the negative goals in the rule bodies are disjoint from the head set and refer only to the parameter set.  With these conditions, the non-membership conditions of negative subgoals will not change during the bottom-up construction, and the modified construction operator will converge.

We start the formalization of this idea by defining rules with negative conditions:

\begin{definition}{Rule with Negative Conditions}
A {\em rule on $U$ with negative conditions} is a triple $(h,B,N)$ where $B, N \subseteq U$ and $h \in U$.  
\end{definition}
For example a Prolog rule with negative conditions, such as $p \mbox{ :- } q, \mbox{ not }r, s, \mbox{ not } t.$ would correspond to the rule $(p, \{q, s\}, \{r, t\})$.

\begin{definition}{The Set of Negative Condition Propositions}
  Given set ${\cal P}$ of rules with negative conditions, the set of negative condition propositions of ${\cal P}$ is:
  \[ N_{\cal P} = \bigcup_{(h,B,N)\in {\cal P}} N \]
\end{definition}

\begin{definition}{Inductive Definition with Negation}
An {\em inductive definition with negation} is a set ${\cal P}$ of rules with negative conditions such that $H_{\cal P} \cap N_{\cal P} = \emptyset$.
\end{definition}
With this constraint, the propositions of $N_{\cal P}$ can get their meanings only from a parameter set.

\begin{definition}{Closed under a Rule with Negation}
A set $S \subseteq U$ is {\em closed under $(h,B,N)$} if $h \in S \mbox{ whenever } B \subseteq S \mbox{ and } N \cap S = \emptyset$.
\end{definition}

\begin{definition}{Set Containing $A$ and Closed under ${\cal P}$ with Negations}
  A set $S$ contains allowable set $A$ and is closed under ${\cal P}$ if
  $A \subseteq S$ and $S$ is closed under every $R \in {\cal P}$.
\end{definition}

\begin{definition}{Set Defined by ${\cal P}$ with Negation and Set $A$}
The set, $F_{\cal P}(A)$, determined by rule set ${\cal P}$ with negation and allowable parameter set $A$ is the least set containing $A$ and closed under ${\cal P}$.  
\end{definition}
If $N_{\cal P} = \emptyset$, this definition reduces to the same definition we gave for positive programs.  Thus, we again use $F_{\cal P}(A)$ for this set.

\begin{figure}
  \includegraphics[width=20cm,height=8cm, trim=2.0cm 4.8cm 0.0cm 2.0cm,
    clip=true]{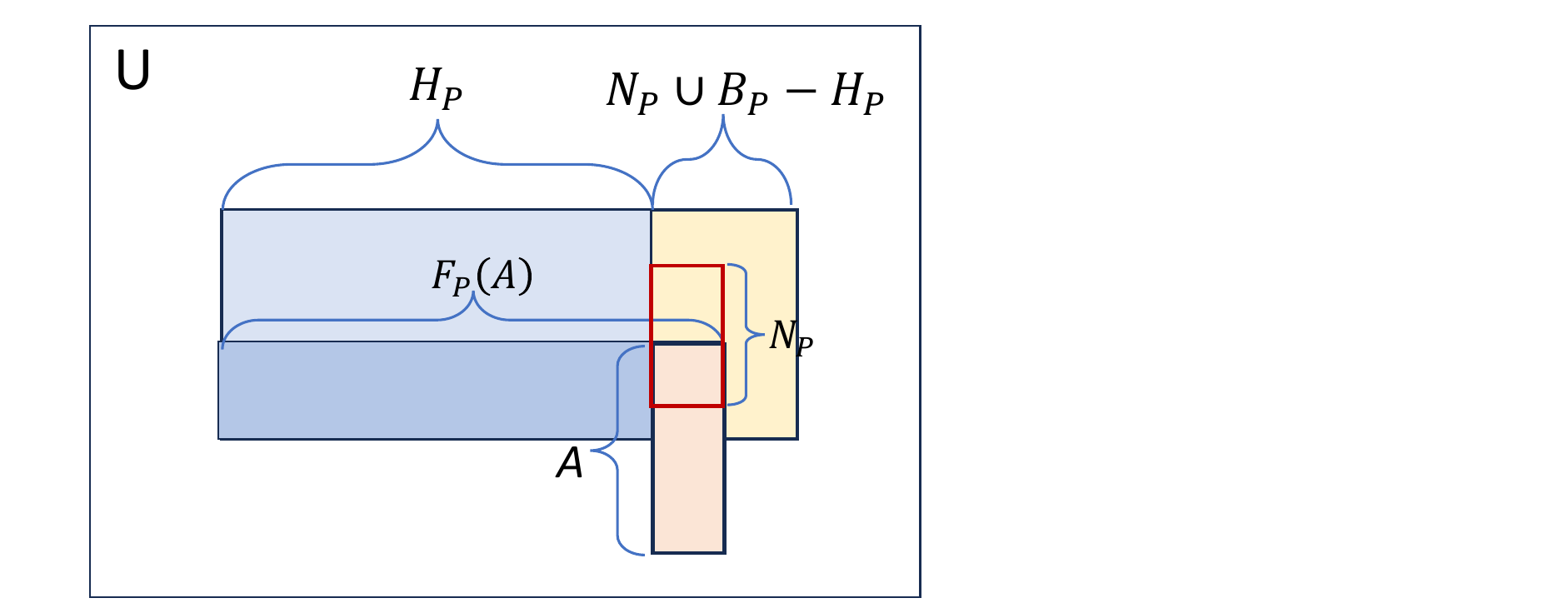}
  \caption{Venn diagram for sets of parameterized ${\cal P}$ with negation}
  \label{theory-venn-neg}
\end{figure}

Figure \ref{theory-venn-neg} shows the structure of a component with negative rules and a parameter set.  Again the large horizontal blue and yellow rectangle represents the propositions of ${\cal P}$, with the blue being the heads, and the yellow being the non-heads.  $N_{\cal P}$, the negated propositions, are outlined by a red box, which is disjoint from $H_{\cal P}$.  The box labeled $A$, a parameter set, is disjoint from $H_{\cal P}$ but intersects with $N_{\cal P}$.  Propositions in that intersection are treated as true, those in $N_{\cal P}$ but not in $A$ are considered as false.  Note that those false propositions will not be in $F_{\cal P}(A)$, since they are not in $H_{\cal P}$.

We next extend the definition of the set function $T_{{\cal P},A}$ to include ${\cal P}$ with rules with negation.
\begin{definition}{The $T_{{\cal P},A}$ Set Function}
  Given set of rules with negation ${\cal P}$ and allowable parameter set $A$,
  \[T_{{\cal P},A}(S) = A \cup \{h : \exists (h,B,N) \in {\cal P} \mbox{ such that } B \subseteq S \wedge N \cap A = \emptyset\}\].
\end{definition}

\begin{definition}{The ${\bf T_{{\cal P},A}^{(i)}}$ Set Operator, Again}
For $A$ allowable for ${\cal P}$, let $T_{{\cal P},A}^{(0)} = \emptyset$; and $T_{{\cal P},A}^{(i+1)}$ = $T_{{\cal P},A}(T_{{\cal P},A}^{(i)})$.  Thus, $T_{{\cal P},A}^{(i)}$ starts with the empty set and iteratively applies $T_{{\cal P},A}$ to it $i$ times.
\end{definition}
This is same definition as for positive programs but now applies to parameterized programs with negation.  And we again get the bottom-up constructive definition, this time for programs with negation:
\begin{theorem} {\normalfont
    $F_{\cal P}(A) = \bigcup_{i=0}^\infty T_{{\cal P},A}^{(i)}$.
} \end{theorem}

A valid decomposition of a program with negation provides a stratification of the propositions of the program.  Propositions that are parameters to a component are in lower strata than propositions in the heads of rules in the component.  The framework here supports a somewhat stronger form of predicate stratification, but not full local stratification.  Predicate stratification requires that the rules can be partitioned using the root symbol of each expression in the head of any rule.  (This implies that a program with a variable as a root head is not predicate stratifiable and not stratifiable by our definition either.)  Our definition requires that heads of parameters not be unifiable with heads of the containing rules.

Recall that Prolog adds the builtin rules:
\begin{verbatim}
  true.
  (A,B) :- A, B.
\end{verbatim}
to describe the semantics of rule bodies.  Prolog programs are expressions and so rules are expressions and so to support metainterpretation, the rules bodies must be given meanings, and that is what these two rules do.  Now we have added a new form for rule bodies, those containing negative literals.  So for example the rule $(p,\{q,t\}, \{r,s\})$ (written in the form in our definitions with a positive set and a negative set) can be written in Prolog as \verb|p :- q, not(r), not(s), t.| (Other orders of the body literals would also be a Prolog form of this rule.)  So negative body literals are represented by trees with root {\tt not/1}.  Thus, we need another rule in the metainterpreter:
\begin{verbatim}
not(X) :- not(X).
\end{verbatim}
This looks like a tautology but is not; the {\tt not} in the head is the root of an expression; the {\tt not} on the right must be treated by the Prolog evaluator as indicating a negative body literal.  I.e., this syntax must be interpreted by the Prolog evaluator as the rule $(not(X),\{\},\{X\})$.  This allows Prolog to metainterpret programs with negative body literals.

Unfortunately, our definition of stratified negation is not strong enough to support this metarule.  Any program containing this rule, indeed this rule itself, is not stratifiable.  Note that {\tt not(a)} is an instance of the head of this rule, and with X instantiated to {\tt not(a)} it is also in the set of negated bodies of instances of this rule.  So there is a nonempty intersection and the instance of this rule cannot be stratified.  Thus, a complete semantics for metaprogramming with Prolog programs with negation must await a treatment of nonstratified negation.  The work Denecker and Vennekens \cite{DBLP:conf/kr/DeneckerV14} on a well-founded semantics for inductive definitions with negation should be directly applicable here and would solve this problem.

\section{First-Order Models of (Meta-)Programs} \label{FOL}
Thus far, we have defined the least propositional model for a set of propositional rules.  Recall that the propositions are, in fact, expressions, i.e., ground atomic formulas in a first-order logic over an appropriate language.  As such, the set of ground atomic formulas uniquely determines a first-order Herbrand model.  We can take that first-order model to characterize the meaning of the given Prolog program.  This is formalized in the definition:
\begin{definition}{First-Order Herbrand Model of ${\cal P}$}
Let $M$ be a first-order Herbrand structure over a super-language of ${\cal P}$.  
Let $M$ be the set of ground atomic formulas true in $M$.  Then
\[ M \models {\cal P} \mbox{ if } M \cap H_{\cal P} = F_{\cal P}((B_{\cal P} - H_{\cal P}) \cap M) \]
\end{definition}
That is, if we use the atomic formulas true in $M$ to determine an allowable parameter set, $A$, then on the heads of ${\cal P}$, $M$ is true for exactly the atomic formulas in the least set containing $A$ and closed under ${\cal P}$.  It is the case that models of the composition of components is the intersection of the models of the components.

Consider an example of a metarule and its meaning in a first-order model:
\begin{verbatim}
truly_believes(X,P) :- believes(X,P), P.
\end{verbatim}
This rule says that if someone believes a proposition {\tt P} and {\tt P} is true, then that someone has a true belief of {\tt P}.  Note that the first two occurrences of {\tt P} in this rule are first-order terms, while the final occurrence is an atomic formula.

Now consider a ground instance of this rule:
\begin{verbatim}
truly_believes(david,tall(marc)) :- believes(david,tall(marc)), tall(marc).
\end{verbatim}
Considered as a first-order statement, the term {\tt tall(marc)} is being used as a name (an object in the domain of the structure) for the atomic formula {\tt tall(marc)} in the logic.  This is a kind of ``Goedel numbering''; the {\em structure} {\tt tall(marc)} serves as a name for the formula {\tt tall(marc)}.  Since the elements of our Herbrand domain, i.e., expressions, look so much like atomic formulas in our logic, Prolog uses this ``identity'' function as its Goedel formula-naming function.

Every term in a Prolog program is the Goedel expression of a ground atomic formula.  Prolog also treats {\tt not({\em X})} as the name for the negation of the ground literal not {\em X}.  (This means, of course, that {\tt not} cannot be a predicate symbol.)

\section{Discussion}

The semantics presented in this paper makes proofs of some program equivalences straightforward.  Consider the transform that takes a Prolog program and wraps {\em every} atomic formula in every rule with a new unary predicate, say {\tt holds/1}.  Prolog programmers know Prolog will give the same answers (modulo the top {\tt hold} wrapper) for the same queries.   Note this works for programs with negation with our semantics, whereas it would not under predicate stratification.  It is easy to see from our semantics that expression {\tt holds({\em E})} is in the meaning of the wrapped program if and only if {\em E} is in the meaning of the first.

As another example, Prolog programmers know that Prolog will give the same result for asking a query $X$ as it does when asking {\tt call({\em X})}.  Why this is true is clear from the definition of {\tt call/1}, i.e. {\tt call(X) :- X}.

Recall our early metainterpreter example:
\begin{verbatim}
true.
(G1,G2) :- G1, G2.
G :- clause(G,Body), Body.
\end{verbatim}
The well-known 3-line Prolog metainterpreter is obtained from this program by wrapping every subgoal (variable or otherwise) with new predicate {\tt interp/1}.  The reasoning for {\tt holds/1} applies here.  The {\tt clause(G,Body)} subgoal is not wrapped for computational reasons, not semantic ones.\footnote{Try it wrapped.  You will find it leads Prolog into an infinite loop.}

We note that a program with a rule with a variable head cannot have a nonempty parameter set and thus cannot be decomposed.  A parameter set must be disjoint from the head set, yet the head set of such a program is all of $U$. The fact that such programs cannot be understood as made up of components may be a good reason for Prolog implementations to disallow them.  And wrapping such rule (sub)sets with a new unary symbol allows the Prolog programmer to effectively get around this limitation when desired.

\section{Why Yet Another Semantics for Prolog?}

Do we need yet another semantics for Prolog?  The traditional minimal Herbrand model semantics for Prolog is based on first-order logic (FOL) and its models and proof theory.  This is appropriate for Prolog as a language for representing data and knowledge since FOL is the accepted foundation for knowledge representation (KR) languages.  But Prolog is also a programming language for implementing algorithms and was clearly understood as such in its early days \cite{DBLP:journals/sigart/WarrenP077}, \cite{DBLP:journals/cacm/Kowalski79}, \cite{DBLP:books/lib/Kowalski79}.  The semantics of this paper is based on simple set induction, a basic computational paradigm.  Induction is fundamentally about computing; FOL is fundamentally about static knowledge representation.

This induction semantics is lower-level, more primitive, more abstract, than the logic-based semantics.  This induction semantics can be used to define the minimal model semantics, as described in Section \ref{FOL}.  And it is clear why: a complete proof system is an inductive definition of the set of logically true expressions, exactly an inductive definition.  But to allow FOL to give an account of Prolog, the non-first-order concept of minimality, or closed world assumption, must be added.  This add-on that is necessary for FOL is intrinsic to inductive definitions.

Also note that another important application of Prolog after KR is grammars.  And grammars are introduced to students in courses on the theory of computation, not as statements in a logic, but as inductive definitions of sets of strings.  Undoubtedly there is a close relationship between logic and grammars (e.g., \cite{DBLP:conf/acl/PereiraW83}), but it seems clear that grammars are fundamentally first understood as inductive definitions.

Teaching the semantics of Prolog to students without a strong background in logic would be much easier with this inductive semantics.  The student need not be taught FO formulas, quantification, FO models, truth in a model, satisfiability, proofs, general substitutions, Skolemization, disjunctive normal form, resolution, Horn clauses, refutation proofs, and finally Herbrand models (in which the previously intuitive FO concept of a function gets specialized into a very non-intuitive particular function).  These are all very important concepts, and they should be understood by students of knowledge representation.  But, they are not required by students who simply want to understand the meaning of a Prolog program as they would understand the meaning of a Lisp program.

And finally, this semantics is compositional, as any self-respecting programming language semantics simply must be.  Programmers understand their programs by understanding their programs' pieces.  A semantics simply {\em must} account for this.  Denecker \cite{DBLP:journals/tocl/DeneckerBM01}, \cite{DBLP:series/lncs/WarrenD23} emphasized the importance of compositionality and we have followed his lead.

\section{Related Work}
There has been much work on metaprogramming in logic programming.  We discuss briefly work deemed relevant to ours.

Costatini \cite{DBLP:conf/birthday/Costantini02} provides a good survey of a number of early approaches to metaprogramming in logic.  Francois Bry \cite{DBLP:journals/tplp/Bry20} explores metalogic in the context of a much larger fragment of logic than just Prolog.

The idea of basing Prolog semantics on a least fixpoint dates back to van Emden and Kowalski's seminal paper \cite{DBLP:journals/jacm/EmdenK76}, in which the $T|_{\cal P}$ operator was defined.  It was further elaborated by Lloyd \cite{DBLP:books/sp/Lloyd84}.  

Our approach to the semantics of Prolog most closely follows that of Denecker et al. \cite{DBLP:journals/tocl/DeneckerBM01,DBLP:journals/corr/abs-2304-13430} in emphasizing parameterized inductive definitions.  Our approach here agrees with Denecker's on Prolog programs that are in first-order logic.  But we also handle metarules (non-first-order rules), treating them as templates for propositional programs.  Also, our treatment of negation is more restrictive than Denecker's, as noted, but we feel ours may have clearer motivations being based on parameterized programs.

\section{Conclusion}

In this paper we have presented a compositional semantics for pure, (finitely locally) stratified Prolog programs and metaprograms.  (We have mentioned how the stratification might be strengthened.)  The semantics given here is a simple generalization of Denecker's semantics of Prolog as inductive definitions (restricted to stratified programs).  By treating rules with variables as templates for propositional rules, we are able to define a first-order model for Prolog meta-programs.  Their meanings agree exactly with what Prolog interpreters do and with Prolog programmers' intuitions.

We want to emphasize the importance of compositionality of Prolog programs, that they can be constructed from components with independent meanings.  Serious programming languages need to be compositional, i.e., programs are made up of independently understandable components.  With a truly non-compositional language, the programmer must keep an entire program in mind when trying to develop a solution to a problem.  And for large programs, that is not possible.  Experientially, Prolog programmers {\em can and do} write very large programs, one piece at a time.  Prolog must be compositional.  But the traditional least model semantics is not compositional; one needs the entire program to define its least model.  This closed world assumption must be applied to the entire program. It is the {\em world} that is being closed.

But in this inductive theory, each component has its own closure assumption, the one that comes with the idea of induction.  Each component has its own ``closed-neighborhood'' assumption.  And each component has its own meaning.  Basing the theory on {\em parameterized} inductive definitions is what makes this possible.  

\section*{Acknowledgements}
I would like especially to thank Marc Denecker for his tireless commitment to interacting with me (and educating me) on important aspects of formal logic and its uses in knowledge representation and programming.
    
%
\bibliographystyle{acmtrans}
\bibliography{metaprogramming_in_prolog}

\end{document}